# The magnetic structure of Li$_2$CuO$_2$: from *ab initio* calculations to macroscopic simulations

*Coen de Graaf,*[(1)] *Ibério de P. R. Moreira,*[(2)] *Francesc Illas,*[(3)]

*Òscar Iglesias*[(4)] *and Amílcar Labarta*[(4)]

(1) Departament de Química Física e Inorgànica and Institut d'Estudis Avançats, Universitat Rovira i Virgili, Plaça Imperial Tàrraco 1, 43005 Tarragona, Spain

(2) Dipartimento di Chimica Inorganica, Chimica Fisica e Chimica dei Materiali, Università di Torino, Via Pietro Giuria 7, I-10125 Torino, Italy

(3) Departament de Química Física and Centre de Recerca en Química Teòrica, Universitat de Barcelona, C/ Martí i Franquès 1, 08028 Barcelona, Spain

(3) Departament de Física Fonamental, Facultat de Física, Universitat de Barcelona, Diagonal 647, 08028 Barcelona, Spain

**Abstract**

The magnetic structure of the edge sharing cuprate compound Li$_2$CuO$_2$ has been investigated by means of ab initio electronic structure calculations. The first and second neighbor in-chain magnetic interactions are calculated to be -142 K and 22 K, respectively. The ratio between the two parameters is smaller than suggested previously in the literature. The interchain interactions are antiferromagnetic in nature and of the order of a few Kelvins only. Monte Carlo simulations using the ab initio parameters to define the model Hamiltonian result in a Néel temperature in rather good agreement with experiment. Spin population analysis situate the magnetic moment on the copper and oxygen ions somewhere between the completely localized picture derived from experiment and the more delocalized picture based on local density calculations.

# I. INTRODUCTION

The impressive richness of the magnetic behavior of the different copper oxide compounds can be traced back to a large extent to the stacking of the $CuO_4$ plaquettes in the lattice. Corner sharing $CuO_4$ units give rise to large antiferromagnetic interactions, while edge sharing units normally result in rather weak ferromagnetic interactions. Depending on the number of linkages between the different $CuO_4$ units, spin chains are formed (neighbors in one direction only) or $CuO_2$ planes appear, typical of the high-$T_c$ superconductors. The combination of edge sharing and corner sharing $CuO_4$ plaquettes can give rise to spin ladders (e.g. the $Sr_{n-1}Cu_nO_{2n-1}$ with $n=2$ series) or zigzag spin chains (e.g. $SrCuO_2$). Based on this geometrical considerations, $Li_2CuO_2$ can be classified as a quasi one-dimensional (1D) spin 1/2 chain formed by edge sharing $CuO_4$ units. Hence, it is expected that the dominant magnetic interaction along the spin chain is ferromagnetic and that there exist additional weaker interchain interactions that account for the non-zero Néel temperature. The sign of the latter interactions cannot be predicted beforehand and must be derived either from interpretation of experimental data or by independent high-level theoretical treatment of the electronic structure.

The magnetic structure of $Li_2CuO_2$ was first described by Sapiña and co-workers.[1] Their neutron scattering experiments indicate that spin ordering sets in at approximately 9 K and consists in an antiferromagnetic (AFM) alignment along the body diagonal of ferromagnetically (FM) ordered spin-chains that run along the *a*-axis (see Fig. 1). The magnetic moment of 0.92 $\mu_B$ was entirely attributed to the $Cu^{2+}$ ion. Later, Boehm and co-workers measured the dispersion of the spin wave excitations in this compound and they interpreted the results with a Heisenberg Hamiltonian in which six different magnetic coupling parameters appear.[2] They classify $Li_2CuO_2$ as an antiferromagnetic insulator with competing magnetic interaction based on their finding that all nearest neighbor interactions, including the in-chain interaction, were predicted to be antiferromagnetic and of similar size. The magnitude of all these interactions was found to be rather small, less than 3 K. In addition, a significant second neighbor in-chain interaction was reported, ferromagnetic in character.

$Li_2CuO_2$ has also been subject of theoretical studies. Several authors performed density functional (DFT) calculations within the local density approximation (LDA) on the periodic structure.[3-7] In all these studies the non-magnetic phase has been found to be metallic and a small band gap of ~0.1 eV is found for the antiferromagnetic alignment of the spin chains.[4,6,7] Weht and Pickett,[4] and Neudert et al.[3] fitted the antibonding band consisting of Cu-$3d_{xy}$ and O-$2p$ orbitals with four different hopping parameters. Both for the in-chain and interchain hopping, the fit results in second neighbor interactions that are larger than the nearest neighbor couplings.[3,4] Moreover, the LDA calculations result in magnetic moments as large as 0.2 $\mu_B$ for the oxygen ions in the compounds, which is claimed to be larger than any experimental O moment.[4] Similar conclusions were derived by Tanaka, Suzuki, and Motizuki,[7] who studied the effect of the introduction of the on-site repulsion in the LDA scheme by applying the LDA+$U$ scheme. For $U$=4 eV, a band gap was found of 0.72 eV. The magnetic moment on oxygen is hardly sensitive to the introduction of the on-site repulsion in the calculation, it only changes from 0.22 $\mu_B$ for LDA to 0.21 $\mu_B$ for LDA+$U$ with $U$=4 eV.

Mizuno et al. analyzed the magnetic interactions in this system by diagonalizing a three band Hubbard Hamiltonian for finite copper-oxide clusters.[8] The model parameters were derived from experiment or taken from the lamellar cuprates $La_2CuO_4$ and $Sr_2CuO_2Cl_2$. The experimental data could be well reproduced by a ferromagnetic nearest neighbor interaction of 100 K and a second neighbor interaction of –60 K, antiferromagnetic in nature. The latter value is reduced to –40 K when interchain interactions along the body diagonal are taken into account. This interchain interaction was calculated to be –16 K.[9]

This surprisingly large second neighbor coupling has been attributed to the short distance between oxygens on the chains which can cause a relatively large overlap between oxygens that connect second neighbor copper ions.[4,8,9] For comparison, the O–O distance in $Li_2CuO_2$ along the chains is 2.86 Å, while the interatomic distance is 3.9 Å for oxygens in corner sharing spin-chain compounds as $Sr_2CuO_3$ and $Ca_2CuO_3$.

In this paper, we apply the well-established computational methods of quantum chemistry as an alternative to the above mentioned approaches to obtain insight in the complex magnetic structure of $Li_2CuO_2$. As an extension of a preliminary study,[10] attention will not only be focused on the accurate determination of the in-chain magnetic parameters, but also on the interchain magnetic interactions and the hopping parameters. The *ab initio* quantum chemical schemes provide a sound hierarchy of increasing accuracy and can be applied both within a periodic and a local (or cluster model) representation of the material. Results obtained over the last decade show that quantum chemical methods, which will be introduced in some more detail in the next Section, are capable to reproduce the nature and the absolute magnitude of magnetic interactions in quantitative agreement with experiment.[11,12] For the present material, experimental data about the magnetic coupling parameters is less clear and the validity of the *ab initio* microscopic electronic structure parameters must be established in a different way. For this purpose, we perform several checks, internal and external to the computational schemes applied. In the first place, we validate the cluster model comparing the results with periodic calculations performed at the same level of approximation. Secondly, the cluster size and basis set dependence of the parameters is investigated. However, the most important check is provided by the determination of several thermodynamic equilibrium quantities through Monte Carlo simulations using the *ab initio* microscopic electronic structure parameters to define the effective magnetic Hamiltonian. These macroscopic quantities can easily be compared with experiment and provide us with a rigorous check on the consistency of the parameters.

## II. QUANTUM CHEMICAL DETERMINATION OF $J$ AND $t$

Figures 2 and 3 illustrate the pathways for the magnetic interactions and hopping processes considered in the present study. In the first place, we focus our attention on the relative magnitude of the in-chain interactions to clarify the uncertainty about the importance of second-neighbor interactions ($J_{b,2}$ and $t_{b,2}$) and the nature of the first neighbor interactions ($J_{b,1}$ and $t_{b,1}$), for which $J_{b,1}$ has been claimed to be antiferromagnetic[1,2] in spite of the almost rectangular nature of the Cu-O-Cu bond and in contradiction to the Goodenough-Kanamori-Anderson rules.[13-15] Secondly, we derive *ab initio* estimates of the interchain interactions. Beside the nearest neighbor interactions along the *a*-axis ($J_{a,1}$ and $t_{a,1}$, not shown in the figures) and the body diagonal ($J_{c,1}$ and $t_{c,1}$), we also consider the next nearest neighbor interaction along the body diagonal ($J_{c,2}$ and $t_{c,2}$). The latter interaction has been claimed to be as important as the nearest neighbor interaction by Mizuno *et al.*[9] Although the copper ions involved in this interaction are more separated than for $J_{c,1}$, the magnetic pathway is identical (Cu-O-Li-O-Cu) for both interactions. From geometrical considerations, it can even be expected that the next nearest neighbor pathway is more favorable (see Figure 3).

### A. Computational methods and material model

Two requisites must be fulfilled for an accurate determination of the electronic structure parameters with a finite representation of the material. In the first place the cluster model must be chosen such that no serious artifacts are introduced. Once the material model is fixed, the *N*-electron eigenfunction of the resulting exact (non-relativistic) cluster Hamiltonian must be approximated in a very accurate way. *Ab initio* cluster model studies performed over the last ten

years established a successful computational strategy to met both criteria.[11,12,16-24]

The cluster model is constructed by including the magnetic centers and its direct neighbors in the quantum cluster region, which is treated at an all-electron level. These atoms are embedded in a set of total ion potentials (TIPs) that represent the cations surrounding the quantum region.[25] Thereafter, optimized point charges are added to account for the long-range electrostatic interactions of the quantum region with the rest of the crystal. The TIPs account for the short-range interaction between cluster atoms and surroundings (Coulomb and exchange interaction) and avoid the spurious delocalization of the charge distribution of the oxygens towards the bare positive point charges. The basic unit to study the in-chain magnetic interactions ($J_{b,1}$ and $J_{b,2}$) is the $Cu_3O_8Li_6$ cluster embedded in two $Cu^{2+}$ TIPs plus point charges. The small number of electrons associated with the $Li^+$ ions permits us to add these ions to the quantum region instead of treating them (more approximately) with TIPs. Similar considerations lead to the following quantum regions for the interchain interactions: $Cu_2O_8Li_4$ for $J_{a,1}$ and $t_{a,1}$; $Cu_2O_8Li_4$ for $J_{c,1}$ and $t_{c,1}$; and $Cu_2O_8Li_2$ for $J_{c,2}$ and $t_{c,2}$. Again, all these cluster models are completed by adding TIPs and optimized point charges. Because no simple relation exist to extract the hopping parameters from a three center cluster,[26] the in-chain hopping parameters $t_{b,1}$ and $t_{b,2}$ are extracted from a $Cu_2O_6Li_4$ and a $Cu_2O_8Li_6$ cluster, respectively. The latter cluster is identical to that used to calculate $J_{b,1}$ and $J_{b,2}$ but for the $Cu^{2+}$ ion in the center of the cluster which is replaced by a 2+ point charge. This modified cluster has been applied before by Mizuno *et al.* to derive $J_{b,2}$ and Sec. 2F will show that the modification does not seriously affect the results.

The Heisenberg Hamiltonian reduces to $\hat{H} = -J\hat{S}_1\hat{S}_2$ for the two center clusters and the magnetic coupling constant is obtained from the energy difference of the singlet and triplet coupled spin states. Positive $J$'s correspond to ferromagnetic interactions and a negative $J$ indicate that antiferromagnetic coupling is preferred. The hopping integral $t$ can be defined as the matrix element of the Hamiltonian between the states in which the hole is localized on center *a* and center *b*. In a symmetry adapted description of the electronic structure, this matrix element corresponds to half the energy difference between the states with the hole in the magnetic orbital of gerade symmetry [ $g = (1/\sqrt{2})(a + b)$ ] and ungerade symmetry [ $u = (1/\sqrt{2})(a - b)$ ].[27,28] The three center cluster allows a simultaneous calculation of $J_{b,1}$ and $J_{b,2}$ using the relations between the spin eigenstates of the Heisenberg Hamiltonian $\hat{H} = -J_1(\hat{S}_1\hat{S}_2 + \hat{S}_2\hat{S}_3) - J_2\hat{S}_1\hat{S}_3$ and the electronic eigenstates of the cluster Hamiltonian. From the mapping we obtain $J_{b,1} = 2/3 [ E(D_u) - E(Q_u) ]$ and $J_{b,2} = J_{b,1} - [ E(D_u) - E(D_g) ]$,[29] where $Q_u$ is the quartet coupled spin state of ungerade symmetry, and $D_u$ and $D_g$ the doublet states of ungerade and gerade symmetry, respectively.

The methods to compute the electronic structure have been applied before to many related transition metal compounds in the study of magnetic coupling constants and hopping parameters. Here, we will only briefly review the main point of the methods, for a more detailed description the reader is referred to previous work. (Refs. [20,21,23] and references therein) The simplest yet physically meaningful approximation of the *N*-electron wave function is a complete active space (CAS) wave function constructed by distributing the unpaired electrons in all possible ways over the magnetic orbitals. This corresponds to the Anderson model of superexchange and will be used here as reference wave function for more elaborate treatments of the electronic structure that include a much larger part of the electron correlation. In the first place, we apply the difference dedicated configuration interaction (DDCI) scheme, which is specially designed to obtain accurate energy differences.[30-32] The method excludes those determinants from the CI wave function that up to second-order perturbation theory do not contribute to the energy difference of the electronic states under study. These are exactly the determinants connected to double replacements from the inactive (or doubly occupied) orbitals into the virtual (or empty) orbitals. Since these determinants are most numerous, the DDCI selection largely reduces the computational cost with almost no loss of accuracy. Moreover, the method has a much smaller size-consistency error than the complete singles-doubles CI.

Because the computational demands are still quite elevated for the DDCI method, we explore the basis set and cluster size dependency of the electronic structure parameters with an alternative method, namely the complete active space second-order perturbation theory (CASPT2).[33,34] This method considers the effect of all single and double replacements but treats them only by second-order perturbation theory. The method has recently been shown to reproduce rather accurately magnetic coupling parameters.[21] Details about the one-electron basis set used to express the atomic orbitals can be found in the Appendix.

## B. Validation of the material model

The most rigorous modelization of a crystal is obtained by imposing periodic boundary conditions on a small building block, typically the unit cell. This way of representing the crystal leads to band structure theory for which various implementations exist. The simplest version is the well-known tight-binding method, which is mainly used for qualitative reasoning. Among the quantitative band structure methods, one of the most popular variants is based on DFT within the local density approximation. The expression of the exchange-correlation part of the functional are based on the non-interacting electron gas. This functional can be improved by adding gradient corrections (GGA methods) or mixing in an arbitrary amount of the exact Fock exchange (the so-called hybrid functionals). Here, we validate our –at first sight somewhat rough– modelization of the crystal by comparing periodic unrestricted Hartree-Fock (UHF) calculations with similar calculations applied to the cluster model. UHF uses the exact non-local exchange, but ignores the electron correlation effects. To a large extent, UHF is the spin unrestricted equivalent of the CASSCF computational scheme mentioned before, i.e. it basically describes the Anderson model, normally results in the correct sign of the interactions, but largely underestimates the experimental values. We apply the linear combination of atomic orbitals (LCAO) approximation to construct the one-electron basis functions in the periodic calculations.

The magnetic coupling parameters are extracted from periodic calculations by comparing the energy per unit cell of different spin alignments.[22,35,36] However, the difference of the FM and AFM spin alignment in the simple unit cell only gives us information about $J_{c,1}$. To obtain estimates for the interactions corresponding to the five closest Cu-Cu distances, we have considered the following double and triple super cells in addition to the simplest one. Doubling along the *a*-axis gives us two different antiferromagnetic spin alignments (AFM2*a*(0) and AFM2*a*(1), the $S_z$ quantum number of the super cell is given in parenthesis) related to $J_{a,1}$ and $J_{a,2}$. Doubling along the *b*-axis gives us two other antiferromagnetic alignments [AFM2*b*(0) and AFM2*b*(1)] and provides a way to extract $J_{b,1}$. Finally, the triplication of the unit cell along the *b*-axis [AFM3*b*(0)] allows us to extract $J_{b,2}$. The calculation of $J_{c,2}$ requires a fourfold super cell and has not been considered because of the very high computational demand. An overview of the computational details of the periodic calculations can be found in the Appendix.

Table I reports the UHF energies per unit cell of the different super cells with respect to the simple FM cell. It also lists the relations between these energies and the magnetic coupling parameters obtained by a mapping onto the Ising Hamiltonian.[37,38] For spin unrestricted calculations, one has to rely on the Ising Hamiltonian because the different spin settings are in general not eigenfunctions of the Heisenberg Hamiltonian.[22,23] Solving the set of linear equations given in Table I results in the following magnetic interaction parameters: $J_{b,1}$ = 127.9 K, $J_{b,2}$ = -5.5 K, $J_{a,1}$ = 0.2 K, $J_{a,2}$ = -0.7 K, and $J_{c,1}$ = -0.2 K. Although the numerical precision of the computational methods applied is better than 0.1 K, the smallness of the interchain interactions makes them less suitable to validate the cluster model. Nevertheless, the in-chain interactions are clearly larger and can be used to make a comparison with the results obtained from a local point of view. Hence, we have used the three center cluster described in the previous section and calculated the UHF energies of the high-spin state [$\alpha(1)\alpha(2)\alpha(3)$, corresponding to a ferromagnetic alignment of the spins on the three copper ions], and two broken symmetry states [$\alpha(1)\alpha(2)\beta(3)$ and $\alpha(1)\beta(2)\alpha(3)$]. From the energy differences, we obtain $J_1$ = 127.0 K and $J_2$ = -6.7 K, in good agreement with the periodic calculations. This comparison validates the modelization of the crystal with an embedded cluster model to extract local electronic structure parameters with more sophisticated quantum chemical schemes than the UHF method used in the periodic calculations. This observation is not unique for $Li_2CuO_2$, but has been reported before for a large series of transition metal oxides and fluorides.[20,22,35,39-42]

The validation of the embedded cluster model approach for the calculation of $t$ cannot be achieved in the same way. Extremely large super cells are needed to obtain a realistic hole concentration to directly calculate the hopping integral in a periodic approach. There exist, however, some indirect support for the suitability of the cluster model approach to calculate $t$'s. In the first place the cluster model satisfactorily reproduces the generally accepted value of $t$ for $La_2CuO_4$.[26] Moreover, the LDA hopping parameters of $Sr_2CuO_3$ and $Ca_2CuO_3$ obtained from cluster model and deduced from periodic LDA calculations[43] are almost identical.[20]

### C. Magnetic interactions

The first cluster model, $Cu_3O_8Li_6$, allows us to calculate both $J_{b,1}$ and $J_{b,2}$. The ferromagnetic character of $J_{b,1}$ found with all three computational methods applied (cf. Table II) is in agreement with the Goodenough-Kanamori-Anderson (GKA) rules.[13-15] Although the Cu-O-Cu angle is not strictly 90º, it is close enough to the ideal situation so that the ferromagnetic contribution to the Cu-Cu interaction is still dominant. To give a more firm basis to the nature of $J_{b,1}$, we have investigated at what angle the antiferromagnetic contribution becomes dominant and the interaction changes sign. To this purpose, we have varied the Cu-O-Cu angle in a $Cu_2O_6Li_4$ cluster model maintaining the Cu-O distance fixed at the experimental value of 1.956 Å, all other cluster atoms and the embedding remained unchanged. The outcome of this computational experiment shows that the nearest neighbor interaction reaches a maximum around 97º and remains ferromagnetic up to angles as large as 104º. For angles smaller than 90º, the interaction becomes antiferromagnetic around 80º. The structure is, however, very stressed at these small angles and the results might be affected by this. Nevertheless, the results shows that the experimental Cu-O-Cu angle of 94º lies in the middle of the ferromagnetic range and, hence, the suggestion of an antiferromagnetic nearest neighbor interaction is not supported.

Comparing the results of the three different computational schemes, we observe the usual behavior. The (almost) uncorrelated CASSCF wave function reproduces the correct sign but the inclusion of the important electron correlation effects by CASPT2 or DDCI largely enhances the interaction. The final result ($J_{b,1}$ = 142 K) is of the same order of magnitude as that derived from the three band model Hamiltonian ($J_{b,1}$ = 100 K),[8,9] but much larger (and of opposite sign) than the one obtained from the fitting of the spin wave dispersions ($J_{b,1}$ = -2.8 K).[2]

The next nearest neighbor in-chain interaction, derived from the same $Cu_3O_8Li_6$ cluster, is antiferromagnetic in nature and, hence, introduces a frustration in the spin chain. The calculated absolute magnitude of the interaction is, however, much smaller than the predictions mentioned in the Introduction. For the CASSCF wave function, $J_{b,2}$ is about 5% of $J_{b,1}$ and can be considered negligible. On the other hand, the explicitly correlated wave functions significantly increase $J_{b,2}$ and our final estimate corresponds to –22 K and a ratio $J_{b,2}$ / $J_{b,1}$ = -0.15. It is interesting to note that the Li ions play an important role in the ratio between $J_{b,1}$ and $J_{b,2}$. When the six lithium ions are removed from the quantum cluster region and represented as bare point charges, $J_{b,2}$ increases dramatically and becomes as large as -102 K. $J_{b,1}$ is much less affected by the removal of the lithium ions and is reduced to 123 K, leading to $J_{b,2} / J_{b,1}$ = -0.83. The large change observed for $J_{b,2}$ indicates that the magnetic interaction path (Cu–O–O–Cu) for this interaction is obstructed when the Li ions around the cluster are represented by real charge distributions instead of point charges. Figure 4 represents the changes in the spin density when Li ions are removed from the quantum cluster region. It clearly illustrates how the introduction of the short-range repulsion between the Li ions and the oxygens on the $J_{b,2}$ magnetic path significantly reduces the spin density along this path to increase it on the Cu ions. Hence, the overlap between the two oxygens decreases and the two copper ions involved in this magnetic interaction are disconnected magnetically.

We now turn to the interchain interactions. The magnetic pathway for these interactions is rather long and complicated (see Figs. 2 and 3) and, therefore, normally result in weak interactions, but they are fundamental to understand the three dimensional magnetic structure of the crystal. The first conclusion that can be drawn from Table II is that the second-order perturbative treatment of the correlation effects is not precise enough for these very small energy differences. The CASPT2 results are much larger than those calculated with the variational DDCI method and result in too high a Néel temperature ($T_N$ ~28 K) when the values are inserted in the mean field expression for $T_N$ of quasi 1D spin chains proposed by Schulz.[44] On the other hand, the DDCI values result in a $T_N$ around 7 K, much closer to the experimental value of 9 K.[1,45,46] Nevertheless, these values have to be taken with caution. In the first place, there is the uncertainty inherent to the mean-field character of the expression,[47-49] and secondly, the five calculated interaction parameters must be converted into one effective in-chain parameter $J_{//}$ and one effective interchain parameter $J_\wedge$. Because $T_N$ is not very sensitive to $J_{//}$ —at least not in the expression of Schulz—, $J_{b,2}$ can either be neglected or the relation $J_{//} = J_{b,1} - rJ_{b,2}$ can be applied with $r = 1$ or $r = 1.12$.[43,50,51] For the interchain interaction, we follow the strategy previously adopted by other authors,[43,52] which consists in taking the average of the interactions perpendicular to the chain as the effective $J_\perp$. For CASPT2, we have used $J_\perp = (J_{a,1} + 1/2(J_{c,1} + J_{c,2}))/2 = -11.8$ K, and for DDCI, $J_\perp = (J_{a,1} + J_{c,1})/2 = -2.5$ K.

In the second place, the DDCI results confirm the assumption of Mizuno *et al*. about the importance of the second neighbor interchain interaction.[9] We obtain, however, a slightly different picture of the interchain interaction along the *c*-axis. Where Mizuno *et al*. assumed that $J_{c,1}$ and $J_{c,2}$ are equal and can be written directly as one effective $J_c$, Table II shows that $J_{c,1}$ is practically zero and $J_{c,2}$ is much larger. In addition, we could determine the strength of the interaction along the *a*-axis, which is approximately half of $J_{c,2}$. We have also investigated the size of $J_{a,2}$, but this interaction turns out to be practically zero with all three computational schemes applied in this work. Therefore, no further reference to this interaction will be made. The relative size of the interaction along the body diagonal ($J_{c,2}$) and in the *a-b* planes ($J_{a,1}$) —both antiferromagnetic in nature— is not incompatible with the experimental magnetic structure, as AFM alignment of the spin chains along the body diagonal is preferred to AFM alignment in the *a-b* planes.

### D. Hopping parameters

The second set of calculations are devoted to the accurate determination of the different *t*'s, which parameterize the dynamics of the holes when the system is doped. The fact that the $CuO_4$ plaquettes are edge sharing has a large effect on the nearest neighbor hopping parameter $t_{b,1}$. Whereas a typical value of this parameter in corner sharing cuprates is around 500 meV, it is more than three times smaller in $Li_2CuO_2$, see Table II. On the other hand, $t_{b,2}$ is of the same order of magnitude as $t_{b,1}$ and almost three times larger than the corresponding *t* in corner sharing cuprates,[26] namely the hopping integral between two copper ions separated by a linear –O–Cu–O– interaction path. The interchain hopping parameters are smaller in magnitude, but not negligible relative to the in-chain parameters. As for the magnetic coupling, we observe that $t_{c,1}$ is significantly smaller than $t_{c,2}$, although the distance between the copper ion is larger for the latter process (5.2 Å versus 6.6 Å). On the contrary, the in-chain hopping parameters are rather similar, unlike the magnetic interactions for which $J_{b,2}$ is only a small fraction of $J_{b,1}$. This seems to indicate that the simple superexchange relation $J = 4t^2 / U$ cannot be applied for $Li_2CuO_2$. Whereas the $J_{b,1}$ and $t_{b,1}$ DDCI values ($U = 4t_{b,1}^2 / J_{b,1}$) result in a reasonable on-site repulsion parameter of 6.7 eV, the DDCI next nearest neighbor interaction parameters lead to an unphysical $U = 26$ eV.

The comparison of the three computational methods applied in this study shows that the CASSCF and DDCI values nearly coincide, whereas the CASPT2 values are significantly larger. The first observation is in agreement with the understanding that the hopping process is basically a one-electron property and therefore not strongly influenced by electron correlation effects. Test calculations in which we only diagonalize a subset of the full DDCI matrix give similar values and confirm the insensitivity of $t$ to electron correlation effects. The second observation indicates that the CASPT2 method is not the best choice to obtain accurate $t$'s. The method also overestimates the hopping parameter for corner sharing cuprates, ~800 meV instead of the usual 500 meV. Nevertheless, CASPT2 perfectly reproduces the trends in the hopping parameters obtained at the more accurate DDCI level. Therefore, it can be perfectly used to explore the basis set and cluster size dependency of the electronic structure parameters presented in Sec. 2F.

**E. Magnetic moments**

The Mulliken spin populations provide a way to extract an estimate of the magnetic moment of the different centers from our cluster calculations. The populations of the CASSCF wave function corresponding to the ferromagnetic solution indicate that a very large part of the magnetic moment is concentrated on the Cu ions. In all clusters, we found that the Mulliken spin population of Cu is 0.93, and ~0.03 for oxygen. Nevertheless, the CASSCF wave function includes only a small amount of electron correlation and more accurate spin populations are needed. Recent work on magnetic moments in molecules learns that DDCI spin populations compare fairly well to experimental results.[53] For $Li_2CuO_2$, we obtain the following DDCI spin populations: 0.76 for copper and 0.12 for oxygen, the spin density on Li is essentially zero. These values are almost independent of the cluster model and the basis set applied. The treatment of the electron correlation effects with DDCI leads to a more delocalized character of the unpaired electrons compared to the CASSCF result, although it does not become as delocalized as found in the LDA calculations.

To give further support to this cluster model results, we have determined the magnetic moments from periodic calculations applying different computational schemes (see the Appendix for computational details). In the first place, there is the already mentioned UHF calculation, which predicts the magnetic moments in excellent agreement with the CASSCF cluster results: 0.90 for Cu and 0.05 for O. Secondly, we performed LDA periodic and cluster calculations. As expected, the periodic LDA calculations give similar results as the previously reported:[4] the spin populations are 0.53 and 0.20 for Cu and O, respectively. These results are accurately reproduced with the LDA cluster model calculation: 0.51 and 0.22 for Cu and O, respectively. Finally, we applied the hybrid B3LYP functional, a gradient corrected functional which has 20% Fock exchange and uses the Lee-Yang-Parr expression for the correlation functional.[54] This functional is one of the most successful functionals in molecular quantum chemistry and has been claimed to reproduce spin densities with reasonable accuracy, although it has the tendency to slightly overestimate the spin density on the bridging ligand.[55,56] Whereas the UHF band gap is unphysically large (16.3 eV) and LDA results in too small a band gap (~ 0.1 eV), the periodic B3LYP calculations give a much more realistic band gap of 2.3 eV, in reasonable agreement with the theoretical estimate reported in the literature.[8] The B3LYP spin densities are 0.65 for copper and 0.17 for oxygen, interpolate between the LDA and UHF results.

Considering the B3LYP values as an upper limit for the oxygen spin density and lower limit for the copper spin density, the results are in good agreement with the DDCI results. We must caution that the way in which the overlap population is divided over the centers —Mulliken population analysis distributes it on equal parts over the two centers involved— is somewhat arbitrary. Nevertheless, it is clear that our results situate the magnetic moments somewhere between the completely localized picture assumed in early experimental work and the more delocalized interpretation based on LDA calculations.

### F. Cluster size and basis set effects

The validation of the calculated electronic structure parameters is continued with a check on the dependence of the $J$'s and $t$'s on the one-electron basis set size. In Table III, we report $J_{b,1}$ and $t_{b,1}$ calculated in the $Cu_2O_6Li_4$ cluster applying five basis sets of different quality. In this series we investigate the effect of a frozen ion description of Li, and the effect of polarization functions on the cluster atoms. The largest basis set considered consists of a ($6s$, $5p$, $4d$, $1f$) basis for Cu; ($5s$, $4p$, $2d$) for O and ($3s$, $1p$) for Li.

The comparison of Basis A, B, D and E shows that the values listed in Table III are converged for the size of the basis set. Adding polarization functions and/or any further extension of the basis set on the cluster ions does not induce significant changes in any of the values. Furthermore, Basis C and D allow to investigate the role of the Li ions, since these basis sets are equivalent except for the description of the Li ions, in the former being as frozen ions not allowing for any covalent interaction with the oxygens. We conclude that a frozen ion description of the Li ions does not seriously affect the magnitude of the magnetic coupling parameters and that the role of the Li ions is (although essential) completely static.

The comparison between periodic and cluster model calculations reported in Sec. 2B have shown that the cluster model provides a valid description of the material to derive microscopic electronic structure parameter. An additional check of the validity of the cluster model can be found in Table IV, where we report the effect of the cluster size on the properties under study. Starting from the $Cu_2O_6Li_4$ cluster used to extract $J_{b,1}$ and $t_{b,1}$, successively more shells are added. The same strategy is applied for the two-center cluster to study the convergence of the second neighbor interactions and the three-center cluster for the simultaneous determination of $J_{b,1}$ and $J_{b,2}$. The largest cluster we consider is $Cu_2O_6Li_{20}O_{16}Cu_2$ (the two extra Cu ions are represented by TIPs) for the two-center cluster and $Cu_xO_8Li_{26}O_{12}Cu_2$ ($x$=2,3) in the second series. Table IV lists the effects of the increase in the cluster size on the magnetic coupling parameters using Basis D for the central cluster atoms ($Cu_2O_6Li_4$ and $Cu_3O_8Li_6$), and (3$s$, 2$p$) and (2$s$) for the other O and Li ions, respectively.

It is readily recognized that the cluster size effect is small, $J_{b,1}$ and $J_{b,2}$ do not significantly depend on the cluster size, provided that the Li ions in the $J_{b,2}$ magnetic pathway are included. Similar considerations apply for the hopping parameters $t_{b,1}$ and $t_{b,2}$. In addition, it can be observed that $J_{b,1}$ derived from the two center cluster is virtually identical to that derived from the three center clusters. Finally, Table IV validates the use of two center clusters to calculate next nearest neighbor interactions. Comparing $Cu_2O_8$ with $Cu_3O_8$, $Cu_2O_8Li_6$ with $Cu_3O_8Li_6$, and so forth, we observe that $J_{b,2}$ is practically identical in both series and that the representation of the central copper by a point charge does not affect the calculated value of $J_{b,2}$. It is assumed that the same applies for $t_{b,2}$.

## III. MONTE CARLO SIMULATIONS

The objectives of the simulations are twofold. In the first place, we determine the Néel temperature $T_N$ for AFM ordering between the FM chains using the *ab initio* magnetic coupling parameters derived in the previous Section. Secondly, we study the dependency of the interchain interactions and the ratio $J_{b,2}/J_{b,1}$ on $T_N$.

### A. Definition of the model

In order to reproduce the crystallographic strcuture of the material and the magnetic interactions between the atoms, we have divided the lattice into two sublattices, each formed by next nearest neighboring *a-b* planes. This allows us to separate the contribution of Cu chains to the equilibrium properties from that of the whole system. Therefore, interplane interactions are represented by interactions between *A* and *B* sublattices. Experimental results[45,57,58] show that there is a strong uniaxial anisotropy along the *a*-axis and, therefore, we have represented the Cu ions by Ising spins $S^{\alpha}_{i,j,k} = \pm 1/2$, where $\alpha$ labels the sublattices *A* and *B* and *i, j, k* represent the vector coordinates in each sublattice. Taking into account the above mentioned comments, the effective model Hamiltonian used in the simulations can be written as

$$\hat{H} = -S^2 \sum_{\alpha=A,B} \sum_{i,j,k=1}^{N} S^{\alpha}_{i,j,k} \left[ J_{b,1} S^{\alpha}_{i,(j+1),k} + J_{b,2} S^{\alpha}_{i,(j+2),k} + J_{a,1} S^{\alpha}_{(i+1),j,k} + J_{c,2} S^{\alpha+1}_{i,j,(k+1)} \right],$$

where *N* is the number of unit cells considered in the simulation and $J_{c,1}$ has been omitted being essentially zero. Moreover, since the *ab initio* calculations indicate that the interchain interactions along the *a*-axis and the *c*-axis are of the same order, we set $J_\perp = (J_{a,1} + J_{c,2})/2$. This reduces the number of parameters in the simulations to two, namely $J_{b,2} / J_{b,1}$ and $J_\perp / J_{b,1}$.

With this Hamiltonian at hand, we have studied several thermodynamic equilibrium quantities through Monte Carlo (MC) simulations. MC techniques has been proven to be a very useful technique for the study of magnetic phase transitions and nature of magnetic order in a wide range of solid-state compounds.[59] It has the advantage that it allows to follow many of the experimentally measured quantities as a function of the temperature or external magnetic fields while keeping track of the microscopic spin configuration not directly accessible by commonly used experimental techniques. In particular, Ising spin lattices with competing ferromagnetic and antiferromagnetic interactions[60] or topologically frustrated lattices[61] have been the object of recent simulation studies aiming at the elucidation of phase diagram of the different possible magnetic order.

We have used periodic boundary conditions and treated systems with linear size up to $N=20$ in order to minimize the finite size effects on the thermodynamical properties. The procedure followed in the MC simulation is the so-called simulated thermal annealing method.[62,63] This method starts with a random spin configuration at very high temperature, which is slowly decreased by a constant factor $\delta T$. We start at a dimensionless temperature $\bar{T}$ of 4 ($\bar{T} = T / J_{b,1} S^2$) and use a reduction factor of –0.005. At each temperature step, the system is brought to thermal equilibrium by evolving the system during a large number of MC steps, normally between 2000 and 5000. The quantities measured after each MC step are the energy $E$, the specific heat $C$, the sublattice magnetizations $M_A$ and $M_B$, and the total magnetization $M_{Total}$.

## B. Simulation results

Figure 5a and inset present the thermal variation of the specific heat $C$ and the energy $E$ during a thermal annealing process for the magnetic coupling parameters derived in Sec. 2C. Setting $J_{b,1} = 1$, ferromagnetic in nature, the simulation parameters are $J_{b,2}/J_{b,1} = -1.549 \cdot 10^{-1}$ and $J_\wedge/J_{b,1} = -1.761 \cdot 10^{-2}$. The sharp peak in $C$ at $\bar{T}_N = 0.61 \pm 0.1$ signals a transition from a paramagnetic phase to antiferromagnetic ordering of the spins. Converting $\bar{T}_N$ in physical units, we obtain $T_N = 10.8 \pm 0.2$ K in good agreement with the experimental value of 9.4 K.[1,45,46] The calculated $T_N$ is stable against a further increase of the system size, no significant changes have been observed for $N > 10$. The nature of the low temperature phase can be understood by looking at the thermal variation of the magnetizations in Fig 5b. The sublattice magnetizations $M_\alpha$ ($\alpha = A, B$) acquire non-zero values at $T_N$ that rapidly saturate to $\pm 1$ at lower $T$. This observation clearly shows that ferromagnetic order in the chains sets in at $T_N$, while the different signs of $M_A$ and $M_B$ indicate that these chains are antiferromagnetically ordered along the $c$-axis. This is completely in agreement with the magnetic structure proposed by Sapiña and co-workers.[1]

To study the effect of the second neighbor in-chain and interchain magnetic interaction parameters on $T_N$, we have run simulations varying the $J_\wedge/J_{b,1}$ ratio from 0 to 0.10 for three different $J_{b,2}/J_{b,1}$ ratios, $-8.000 \cdot 10^{-2}$, $-1.549 \cdot 10^{-1}$ and $-2.500 \cdot 10^{-1}$. Result are given in Fig. 6. In the first place, we observe that $T_N$ vanishes below a certain value of $J_\wedge/J_{b,1}$, indicating that a finite value of the interchain interaction is neccesary to induce AF order. As expected from the AF nature of the interchain interactions, $T_N$ increases with increasing interchain interaction. On the other hand, the increase in $J_{b,2}$ result in a decrease of the Néel temperature because of the increasing frustration in the spin chain.

## IV. SUMMARY AND CONCLUSIONS

An extended *t-J* Hamiltonian of the quasi 1D spin chain compound $Li_2CuO_2$ have been parametrized by means of state-of-the-art *ab initio* quantum chemistry calculations. We have established the ferromagnetic nature of the first neighbor in-chain magnetic interaction (142 K) and observe that the second neighbor in-chain magnetic interaction is antiferromagnetic in nature and about 15% of the first neighbor interaction, and we conclude that the frustration in the spin chain is significantly smaller than suggested by others. At first sight this could be incompatible with the small $T_N$ observed for $Li_2CuO_2$. A way to assess the 3D magnetic structure of the compound (more especifically, $T_N$) was opened by completing the model Hamiltonian with interchain interactions. These antiferromagnetic interactions are very weak (-3.6 K for the interaction along the c-axis and –1.4 along the a-axis) and suggest a very low AF ordering temperature $T_N$. The hopping parameters show a very similar pattern, with the exception of the ratio between the first and second neighbor in-chain hopping parameters, which is much larger than the ratio of the corresponding magnetic interactions.

The validity of the parameters have been checked with three different approaches. In the first place we compare our cluster model results with band structure calculations. The comparison at the UHF level shows that the magnetic interactions parameters are essentially identical in the two representations of the model, e.g. the $J_{b,2} / J_{b,1}$ ratio obtained in the periodic UHF calculations is 0.04, in very good agreement with the UHF cluster model result. This validates our modeling of the material with a finite cluster model. In the second place, we study the cluster size and basis set dependency of the electronic structure parameters. Neither for the cluster size nor for the basis set, we observe significant changes, once a reasonable choice have been made. Finally, and most importantly, we use our *ab initio* parameters to define an effective model Hamiltonian that permits us to perform Monte Carlo simulations of the system. The resulting Néel temperature of

10.8 K is in good agreement with the experimental value, showing that a small $J_{b,2}/J_{b,1}$ ratio does not neccesarily lead to high ordering temperatures. Moreover, the Monte Carlo simulations suggest that the system is rather close to a situation for which no longer three dimensional magnetic ordering occurs.

**REFERENCES**

1. F. Sapiña, J. Rodríguez-Carvajal, M. J. Sanchis, R. Ibáñez, A. Beltrán and D. Beltrán, Solid State Comm. **74**, 779 (1990).

2. M. Boehm, S. Coad, B. Roessli, A. Zheludev, M. Zolliker, P. Böni, D. M. Paul, H. Eisaki, N. Motoyama and S. Uchida, Europhys. Lett. **43**, 77 (1998).

3. R. Neudert, H. Rosner, S.-L. Drechsler, M. Kielwein, M. Sing, Z. Hu, M. Knupfer, M. S. Golden, J. Fink, N. Nücker, M. Merz, S. Schuppler, N. Motoyama, H. Eisaki, S. Uchida, M. Domke and G. Kaindl, Phys. Rev. B **60**, 13413 (1999).

4. R. Weht and W. E. Pickett, Phys. Rev. Lett. **81**, 2502 (1998).

5. H. Rosner, R. Hayn and S.-L. Drechsler, Physica B **259-261**, 1001 (1999).

6. N. Tanaka, M. Suzuki and K. Motizuki, J. Phys. Soc. Jap. **68**, 1684 (1999).

7. N. Tanaka, M. Suzuki and K. Motizuki, Physica B **284-288**, 1388 (2000).

8. Y. Mizuno, T. Tohyama, S. Maekawa, T. Osafune, N. Motoyama, H. Eisaki and S. Uchida, Phys. Rev. B **57**, 5326 (1998).

9. Y. Mizuno, T. Tohyama and S. Maekawa, Phys. Rev. B **60**, 6230 (1999).

10. C. de Graaf, I. de P. R. Moreira and F. Illas, Int. J. Mol. Sci. **1**, 28 (2000).

11. I. de P. R. Moreira, F. Illas, C. J. Calzado, J. F. Sanz, J.-P. Malrieu, N. Ben Amor and D. Maynau, Phys. Rev. B **59**, 6593 (1999).

12. D. Muñoz, F. Illas and I. de P. R. Moreira, Phys. Rev. Lett. **84**, 1579 (2000).

13. J. B. Goodenough, Phys. Rev. **100**, 564 (1955).

14. J. Kanamori, J. Phys. Chem. Solids **10**, 87 (1959).

15. P. W. Anderson, *Theory of magnetic exchange interaction: Exchange in insulators and semiconductors*, Solid State Physics, Vol. 14 (Academic Press, New York, 1963).

16. F. Illas, J. Casanovas, M. A. Garcia-Bach, R. Caballol and O. Castell, Phys. Rev. Lett. **71**,

3549 (1993).

[17] J. Casanovas, J. Rubio and F. Illas, Phys. Rev. B **53**, 945 (1996).

[18] C. de Graaf, R. Broer and W. C. Nieuwpoort, Chem. Phys. Lett. **271**, 372 (1997).

[19] C. Wang, K. Fink and V. Staemmler, Chem. Phys. **192**, 25 (1995).

[20] C. de Graaf and F. Illas, Phys. Rev. B **63**, 014404 (2001).

[21] C. de Graaf, C. Sousa, I. de P. R. Moreira and F. Illas, J. Phys. Chem. A **105**, 11371 (2001).

[22] I. de P. R. Moreira and F. Illas, Phys. Rev. B **55**, 4129 (1997).

[23] F. Illas, I. de P. R. Moreira, C. de Graaf and V. Barone, Theor. Chem. Acc. **104**, 265 (2000).

[24] A. B. van Oosten, R. Broer and W. C. Nieuwpoort, Chem. Phys. Lett. **257**, 207 (1996).

[25] N. W. Winter, R. M. Pitzer and D. K. Temple, J. Chem. Phys. **86**, 3549 (1987).

[26] C. J. Calzado and J.-P. Malrieu, Phys. Rev. B **63**, 214520 (2001).

[27] C. J. Calzado, J. F. Sanz and J.-P. Malrieu, J. Chem. Phys. **112**, 5158 (2000).

[28] J. F. Sanz and J.-P. Malrieu, J. Phys. Chem. **97**, 99 (1993).

[29] E. Sinn, Coor. Chem. Rev. **5**, 313 (1970).

[30] J. Miralles, J.-P. Daudey and R. Caballol, Chem. Phys. Lett. **198**, 555 (1992).

[31] J. Miralles, O. Castell, R. Caballol and J.-P. Malrieu, Chem. Phys. **172**, 33 (1993).

[32] V. M. García, O. Castell, R. Caballol and J.-P. Malrieu, Chem. Phys. Lett. **238**, 222 (1995).

[33] K. Andersson, P.-Å. Malmqvist, B. O. Roos, A. J. Sadlej and K. Wolinski, J. Phys. Chem. **94**, 5483 (1990).

[34] K. Andersson, P.-Å. Malmqvist and B. O. Roos, J. Chem. Phys. **96**, 1218 (1992).

[35] J. M. Ricart, R. Dovesi, C. Roetti and V. R. Saunders, Phys. Rev. B **52**, 2381 (1995).

[36] M. D. Towler, N. L. Allan, N. M. Harrison, V. R. Saunders, W. C. Mackrodt and E. Aprà, Phys. Rev. B **50**, 5041 (1994).

[37] L. Noodleman and J. G. Norman Jr., J. Chem. Phys. **70**, 4903 (1979).

[38] L. Noodleman and E. R. Davidson, Chem. Phys. **109**, 131 (1986).

[39] Y.-S. Su, T. A. Kaplan, S. D. Mahanti and J. F. Harrison, Phys. Rev. B **59**, 10521 (1999).

[40] P. Reinhardt, M. P. Habas, R. Dovesi, I. de P. R. Moreira and F. Illas, Phys. Rev. B **59**, 1016 (1999).

[41] P. Reinhardt, I. de P. R. Moreira, C. de Graaf, R. Dovesi and F. Illas, Chem. Phys. Lett. **319**, 625 (2000).

[42] I. de P. R. Moreira and F. Illas, Phys. Rev. B **60**, 5179 (1999).

[43] H. Rosner, H. Eschrig, R. Hayn, S.-L. Drechsler and J. Málek, Phys. Rev. B **56**, 3402 (1997).

[44] H. J. Schulz, Phys. Rev. Lett. **77**, 2790 (1996).

[45] K. Okuda, S. Noguchi, K. Konishi, H. Deguchi and K. Takeda, J. Magn. Magn. Mat. **104-107**, 817 (1992).

[46] S. Ebisu, T. Komatsu, N. Wada, T. Hashiguchi, P. Kichambare and S. Nagata, J. Phys. Chem. Solids **59**, 1407 (1998).

[47] S. G. Chung and Y. C. Chang, J. Phys. A **20**, 2875 (1987).

[48] V. Y. Irkhin and A. A. Katanin, Phys. Rev. B **61**, 6757 (2000).

[49] A. W. Sandvik, Phys. Rev. Lett. **83**, 3069 (1999).

[50] A. Fledderjohann and C. Gros, Europhys. Lett. **37**, 189 (1997).

[51] D. Gottlieb, M. Lagos, K. Hallberg and C. Balseiro, Phys. Rev. B **43**, 13668 (1991).

[52] A. B. van Oosten and F. Mila, Chem. Phys. Lett. **295**, 359 (1998).

[53] J. Cabrero and R. Caballol, unpublished

[54] A. D. Becke, J. Chem. Phys. **98**, 5648 (1993).

[55] E. Ruiz, J. Cano, S. Alvarez and P. Alemany, J. Am. Chem. Soc. **120**, 11122 (1998).

[56] C. Blanchet-Boiteux and J.-M. Mouesca, J. Am. Chem. Soc. **122**, 861 (2000).

[57] H. Ohta, N. Yamauchi, T. Nanba, M. Motokawa, S. Kawamata and K. Okuda, J. Phys. Soc.

Jap. **62**, 785 (1993).

[58] R. J. Ortega, P. J. Jensen, K. V. Rao, F. Sapiña, D. Beltrán, Z. Iqbal, J. C. Cooley and J. L. Smith, J. Appl. Phys. **83**, 6542 (1998).

[59] K. Binder and D. W. Heermann, *Monte Carlo methods in Statistical Physics*, Springer Series in Solid State Sciences, Vol. 80 (Springer-Verlag, Berlin, 1988).

[60] O. Iglesias and A. Labarta, Phys. Rev. B **63**, 184416 (2001).

[61] S. T. Bramwell and M.-P. Gingras, Science **294**, 1495 (2001).

[62] O. Iglesias and A. Labarta, J. Magn. Magn. Mat. **221**, 149 (2000).

[63] S. Kirpatrick, D. C. Gellat and M. P. Vecchi, Science **220**, 671 (1983).

[64] R. Pou-Amérigo, M. Merchán, I. Nebot-Gil, P.-O. Widmark and B. O. Roos, Theor. Chim. Acta **92**, 149 (1995).

[65] P.-O. Widmark, P.-Å. Malmqvist and B. O. Roos, Theor. Chim. Acta **77**, 291 (1990).

[66] D. Maynau and N. Ben Amor, CASDI suite of programs, Université Paul Sabatier, Toulouse, 1997

[67] K. Andersson, M. Barysz, A. Bernhardsson, M. R. A. Blomberg, D. L. Cooper, T. Fleig, M. P. Fülscher, C. de Graaf, B. A. Hess, G. Karlström, R. Lindh, P.-Å. Malmqvist, P. Neogrády, J. Olsen, B. O. Roos, B. Schimmelpfennig, M. Schütz, L. Seijo, L. Serrano-Andrés, P. E. M. Siegbahn, J. Stålring, T. Thorsteinsson, V. Veryazov and P.-O. Widmark, MOLCAS Version 5, Lund University, Sweden, 2000.

[68] M. J. Frisch, G. W. Trucks, H. B. Schlegel, G. E. Scuseria, M. A. Robb, J. R. Cheeseman, V. G. Zakrzewski, J. A. Montgomery Jr., R. E. Stratmann, J. C. Burant, S. Dapprich, J. M. Millam, A. D. Daniels, K. N. Kudin, M. C. Strain, O. Farkas, J. Tomasi, V. Barone, M. Cossi, R. Cammi, B. Mennucci, C. Pomelli, C. Adamo, S. Clifford, J. Ochterski, G. A. Petersson, P. Y. Ayala, Q. Cui, K. Morokuma, D. K. Malick, A. D. Rabuck, K. Raghavachari, J. B.

Foresman, J. Cioslowski, J. V. Ortiz, A. G. Baboul, B. B. Stefanov, G. Liu, A. Liashenko, P. Piskorz, I. Komaromi, R. Gomperts, R. L. Martin, D. J. Fox, T. Keith, M. A. Al-Laham, C. Y. Peng, A. Nanayakkara, C. Gonzalez, M. Challacombe, P. M. W. Gill, B. Johnson, W. Chen, M. W. Wong, J. L. Andres, C. Gonzalez, M. Head-Gordon, E. S. Replogle and J. A. Pople, Gaussian 98 (Revision A.7), Gaussian, Inc., Pittsburgh, PA, 1998.

69  V. R. Saunders, R. Dovesi, C. Roetti, M. Causà, N. M. Harrison, R. Orlando and C. M. Zicovich-Wilson, CRYSTAL98 User's Manual, University of Torino, Torino, 1998.

70  Atomic basis sets are taken from the CRYSTAL web site: http://www.ch.unito.it/ifm/teorica/Basis_Sets/mendel.html and http://www.ch.unito.it/ifm/teorica/AuxB_Sets/mendel.html

**APPENDIX**

The results listed in Table II have been obtained using one-electron basis sets of the atomic natural orbital type. The Cu basis set is a general contraction of the ($21s$, $15p$, $10d$) primitive set to [$5s$, $4p$, $3d$] gaussian type functions. We use a ($14s$, $9p$) / [$4s$, $3p$] basis set for O and a ($14s$) / [$2s$] basis set for Li.[64,65] This corresponds to basis B in Table III and results in 212 basis functions for the $Cu_3O_8Li_6$ cluster model. Both in the DDCI and the CASPT2 calculations, the deep-core electrons (Cu $1s^2$, $2s^2$, $2p^6$ and O $1s^2$) were kept frozen. The DDCI and CASPT2 calculations have been performed with the CASDI and MOLCAS 5.2 codes,[66,67] respectively. The LDA and UHF cluster model calculations have been perfomed with the GAUSSIAN98 program,[68] applying the following segmented basis sets: 6-3111+G for Cu, 6-31G* for O and STO-3G for Li.

The CRYSTAL98 program[69] has been used for all periodic electronic structure calcualtions reported here. Standard basis sets have been used in the periodic calculations,[70] i.e. ($20s$, $12p$, $5d$) / [$5s$, $4p$, $2d$] for Cu, ($14s$, $6p$) / [$4s$, $3p$] for O, and ($7s$, $1p$) / ($2s$, $1p$) for Li, where a segmented contraction scheme is applied. The cutoff parameters for the Coulomb and exchange integral evaluation (ITOL 1-5 of the CRYSTAL98 code) have been set to 7, 7, 7, 7, 14. The $k$-space grid parameter is 6 for the double super cells and 4 for the triple super cells, yielding 67 and 27 $k$-points in the first irreducible Brillouin zone, respectively. This parameter choice is taken from previous applications[20,35,39] and results in an energy difference per cell smaller than $10^{-6}$ hartree for the FM alignment in the single and triple unit cells and even better for the difference between single and double unit cells.

TABLE I. UHF relative energies per formula unit of different spin settings in simple (AFM and FM), double (AFM2$a$(0), AFM2$a$(1), AFM2$b$(1) and AFM2$b$(0)) and triple (AFM3$b$(0)) cells. The relations resulting from a mapping onto the Ising Hamiltonian are also given.

| Super cell | Relation | Relative energy (in K) |
|---|---|---|
| AFM2$a$(0) | $4J_{a,2} + 4J_{c,1}$ | -3.474 |
| AFM2$a$(1) | $J_{a,1} + 2J_{a,2} + 4J_{c,1}$ | -1.895 |
| AFM | $8J_{c,1}$ | -1.263 |
| FM | 0 | 0.000 |
| AFM2$b$(1) | $J_{b,1} + 2J_{a,2} + 4J_{c,1}$ | 125.994 |
| AFM3$b$(0) | $2J_{b,1} + 4J_{a,2} + 4J_{c,1} + 2J_{b,2}$ | 241.412 |
| AFM2$b$(0) | $2J_{b,1} + 4J_{a,2} + 4J_{c,1}$ | 252.303 |

TABLE II. Magnetic coupling parameters (in K) and hopping integrals (in meV) for $Li_2CuO_2$. CASSCF represents the Anderson model, whereas CASPT2 and DDCI include external electron correlation effects. $J_{b,1}$ and $J_{b,2}$ parameterize the nearest and next nearest neighbor in-chain magnetic interaction, respectively. $J_{a,1}$ stands for the interchain interaction along the *a*-axis. $J_{c,1}$ and $J_{c,2}$ are the nearest and next nearest neighbor interchain interactions along the *c*-axis. For *t*, analogous nomenclature is applied. See also Figs. 2 and 3.

| Method | $J_{b,1}$ | $J_{b,2}$ | $J_{a,1}$ | $J_{c,1}$ | $J_{c,2}$ |
|---|---|---|---|---|---|
| CASSCF | 45 | -3 | 0.0 | 0.0 | -0.4 |
| CASPT2 | 150 | -21 | -10.2 | -12.8 | -13.9 |
| DDCI | 142 | -22 | -1.4 | 0.0 | -3.6 |
|  | $t_{b,1}$ | $t_{b,2}$ | $t_{a,1}$ | $t_{c,1}$ | $t_{c,2}$ |
| CASSCF | 179 | 125 | -9 | 28 | -53 |
| CASPT2 | 322 | 267 | -12 | 67 | -115 |
| DDCI | 143 | 120 | -8 | 28 | -52 |

TABLE III. Basis set dependency of the magnetic interaction parameter $J_{b,1}$ (in K) and the hopping parameter $t_{b,1}$ (in meV) for $Li_2CuO_2$ calculated with an embedded $Cu_2O_6Li_4$ cluster. Basis A consists of the Cu (5s, 4p, 3d) basis, the bridging O (4s, 3p) basis, the edge O (3s, 2p) basis and a Li (2s) basis. Basis B augments the edge O basis to (4s, 3p). Basis C augments B with a d-function on all O but treats the Li ions at the frozen ion level. Basis D only differs from C in the treatment of the Li ions, namely by a (3s, 1p) basis. Basis E consists of a (6s, 5p, 4d, 1f) basis for Cu, a (5s, 4p, 2d) basis for O and a (3s, 1p) basis for Li.

|  | $J_{b,1}$ | | $t_{b,1}$ | |
| --- | --- | --- | --- | --- |
|  | CASSCF | CASPT2 | CASSCF | CASPT2 |
| Basis A | 44 | 147 | 180 | 320 |
| Basis B | 45 | 150 | 179 | 322 |
| Basis C | 46 | 159 | 165 | 316 |
| Basis D | 45 | 150 | 164 | 330 |
| Basis E | 44 | 147 | 160 | 356 |

TABLE IV. Cluster size dependency of the CASPT2 in-chain magnetic coupling parameters $J_{b,1}$ and $J_{b,2}$ (in K), and the in-chain hopping integrals $t_{b,1}$ and $t_{b,2}$ (in meV). All clusters are embedded in two $Cu^{2+}$ TIPs and point charges.

| Cluster | $J_{b,1}$ | $J_{b,2}$ | $t_{b,1}$ | $t_{b,2}$ |
|---|---|---|---|---|
| $Cu_2O_6$ | 132 | – | 329 | – |
| $Cu_2O_6Li_4$ | 150 | – | 330 | – |
| $Cu_2O_6Li_{20}$ | 153 | – | 320 | – |
| $Cu_2O_6Li_{20}O_{16}$ | 156 | – | 316 | – |
| $Cu_2O_8$ | – | -93 | – | 486 |
| $Cu_2O_8Li_6$ | – | -37 | – | 278 |
| $Cu_2O_8Li_{26}$ | – | -30 | – | 295 |
| $Cu_2O_8Li_{26}O_{12}$ | – | -33 | – | 230 |
| $Cu_3O_8$ | 138 | -99 | – | – |
| $Cu_3O_8Li_6$ | 153 | -22 | – | – |
| $Cu_3O_8Li_{10}$ | 154 | -22 | – | – |
| $Cu_3O_8Li_{26}$ | 167 | -27 | – | – |
| $Cu_3O_8Li_{26}O_{12}$ | 163 | -25 | – | – |

FIGURE CAPTIONS:

FIG 1. Crystal structure of the quasi-1D spin-chain $Li_2CuO_2$. Small dark gray spheres represent copper ions, large gray spheres depict the oxygen ions and the light spheres the lithium ions.

FIG 2. Interaction pathways (marked by black lines) for the in-chain magnetic coupling and hopping parameters.

FIG 3. Interaction pathways (marked by black lines) for the interchain magnetic coupling and hopping parameters between chains located in different *a-b* planes. For the nearest neighbor interactions ($J_{c,1}$ and $t_{c,1}$) six equivalent pathways can be defined.

FIG 4. Changes in the CASSCF spin density on the addition of the Li ions to the quantum cluster region. Solid contours indicate a decrease of the spin density, whereas the dotted contours enclose areas of increasing spin density.

FIG 5. (a) Thermal variation of the specific heat *C* as obtained from MC simulation applying the ab initio calculated *J*'s. The position of the peak in C marks the Néel temperature $\bar{T}_N$ for antiferromagnetic ordering. Inset: thermal dependence of the magnetic energy *E*. (b) Thermal variation of sublattice magnetizations $M_A$ and $M_B$, and total magnetization of the system $M_{Total}$. $M_\alpha = \pm 1$ corresponds to complete FM order along the chains.

FIG 6. Dimensionless Néel temperature $\bar{T}_N$ as function of the ratio $J_\perp / J_{b,1}$ for three $J_{b,2}$ to $J_{b,1}$ ratios. Circles correspond to $J_{b,2} / J_{b,1} = -8.000 \cdot 10^{-2}$, squares to $J_{b,2} / J_{b,1} = -1.549 \cdot 10^{-1}$ and triangles give $J_{b,2} / J_{b,1} = -2.500 \cdot 10^{-1}$. $\bar{T}_N$ corresponding to the *ab initio* values derived in Sec. 2C is marked by an empty circle.

Figure 1

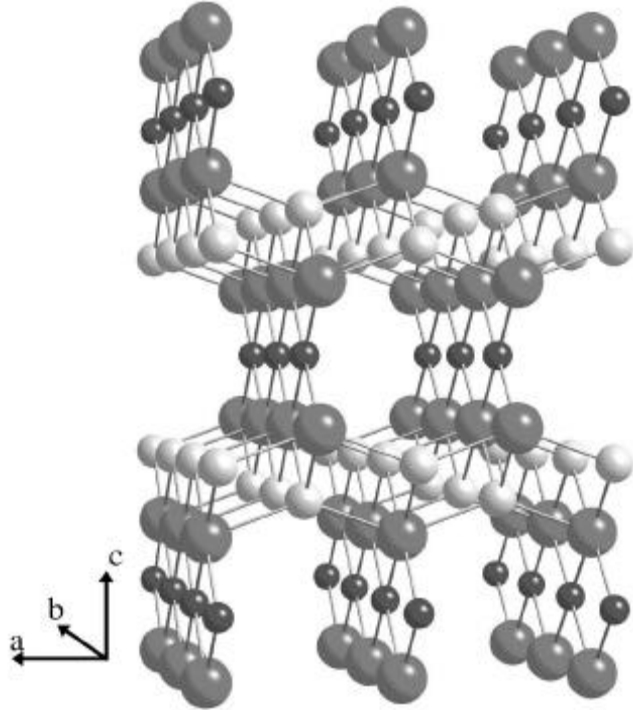

Figure 2

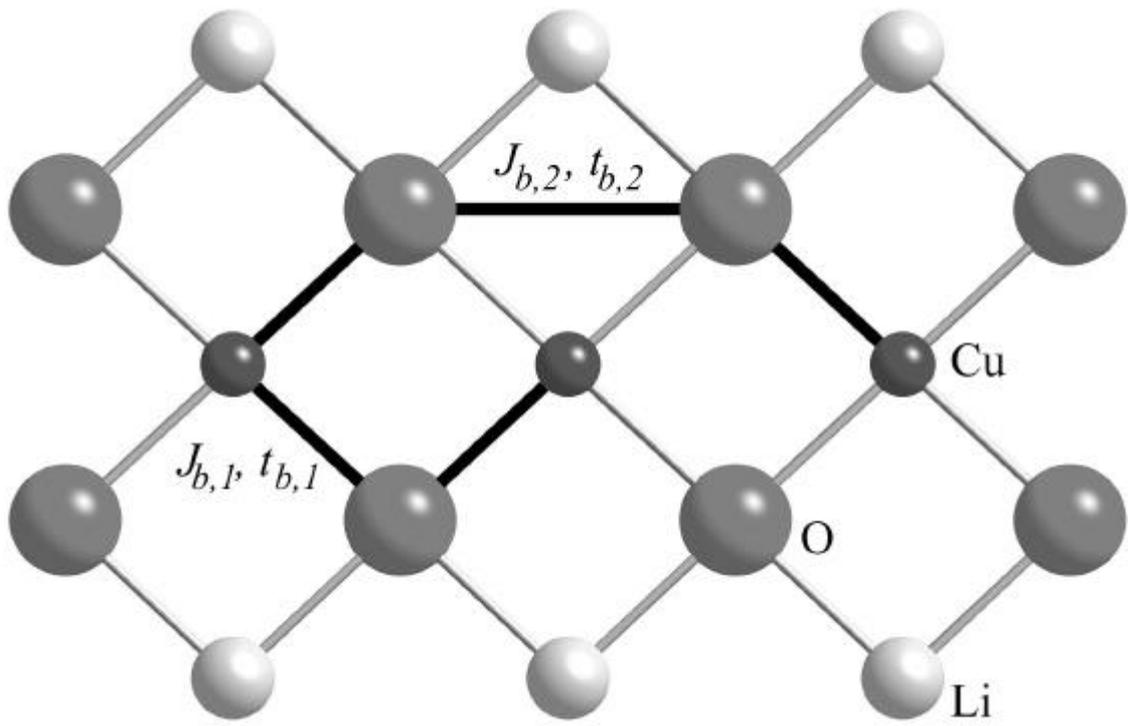

Figure 3.

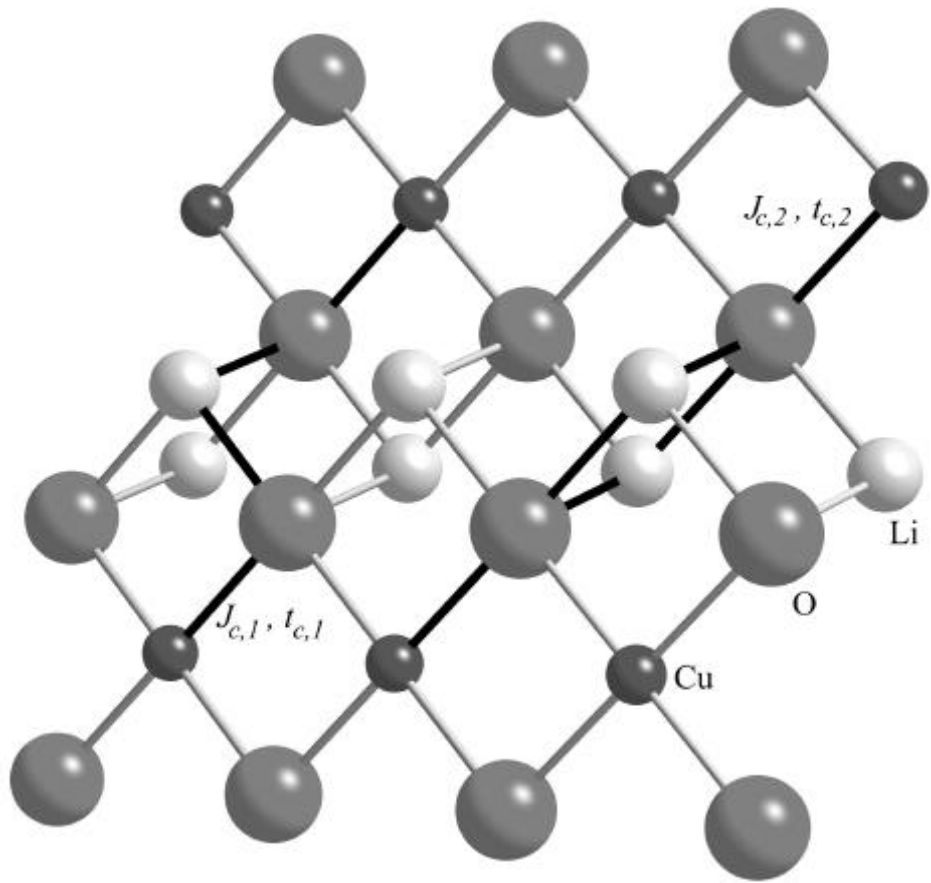

Figure 4.

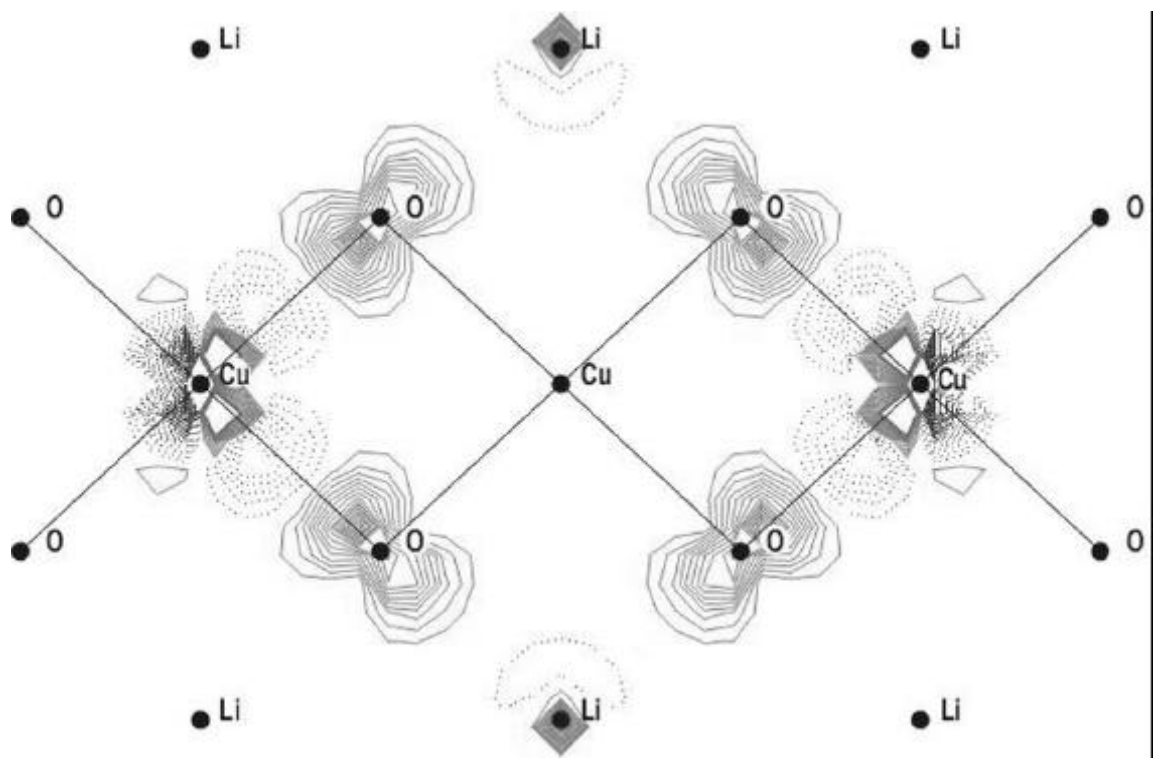

Figure 5.

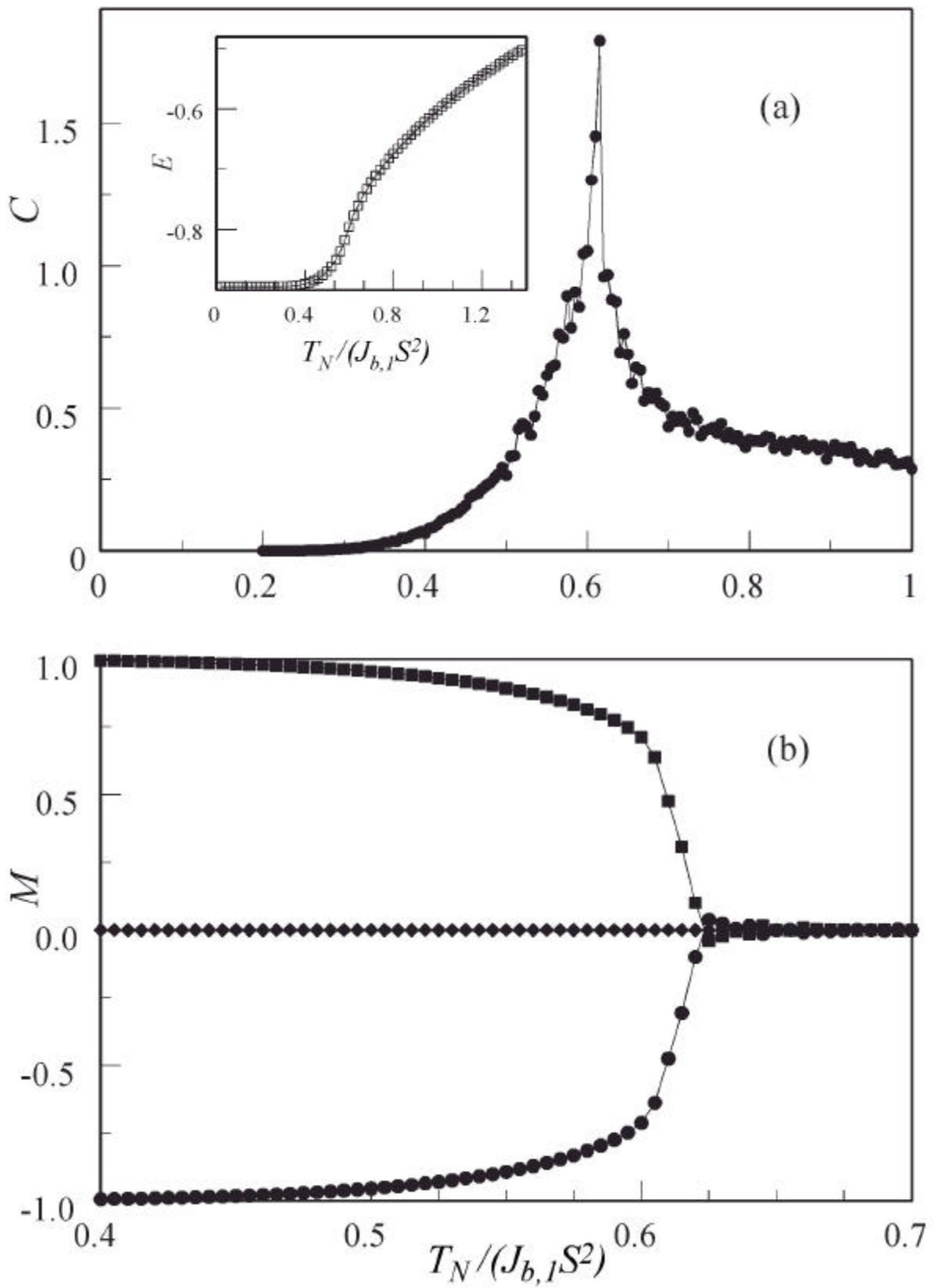

Figure 6.

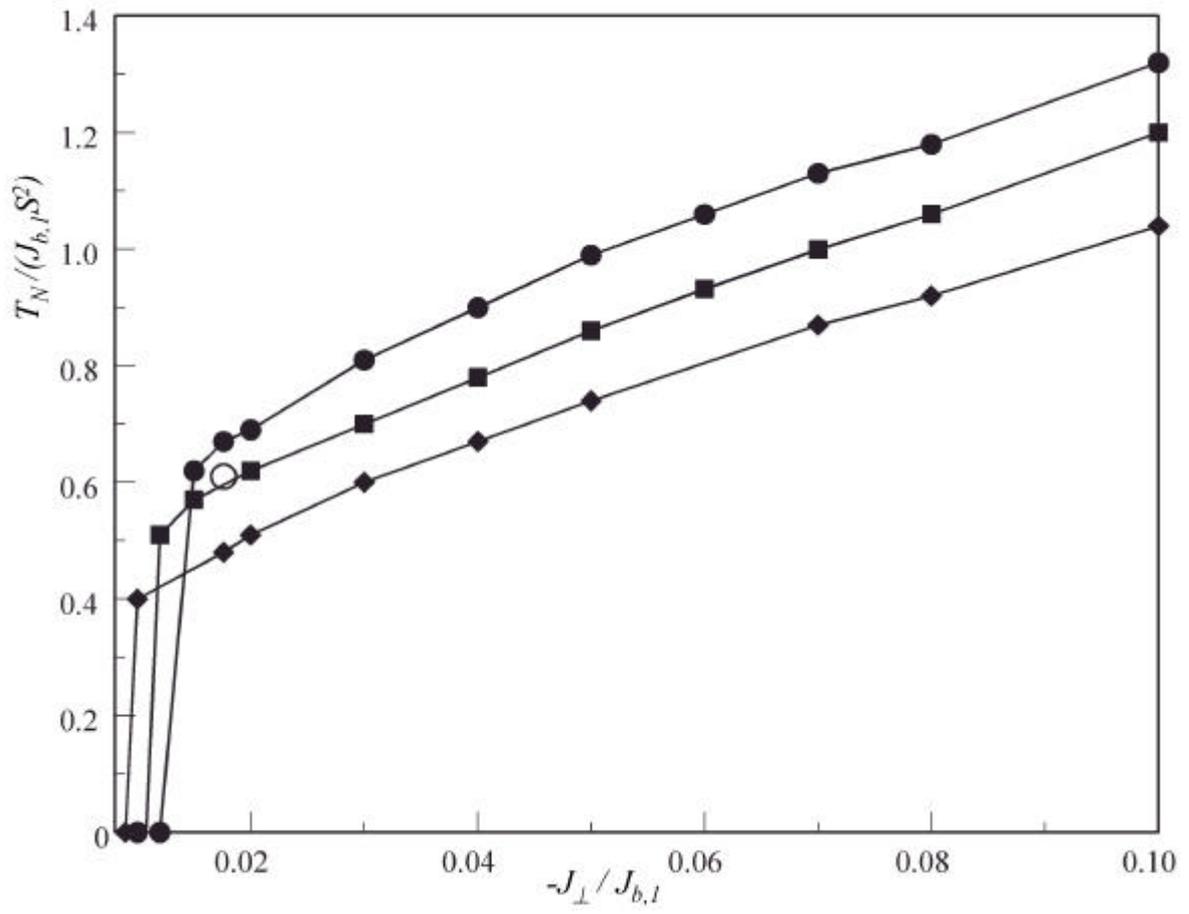